\documentclass[conference]{IEEEtran}
\IEEEoverridecommandlockouts
\usepackage{cite}
\usepackage{booktabs}
\usepackage{multirow}

\usepackage{amsmath,amssymb,amsfonts}
\usepackage{algorithmic}
\usepackage[linesnumbered,ruled,vlined]{algorithm2e}   
 
\usepackage{graphicx}
\usepackage{textcomp}
\usepackage{xcolor}
\usepackage{enumitem}
\usepackage{seqsplit}
\usepackage{subcaption}
\usepackage{makecell}
\usepackage{stfloats}
\usepackage[most]{tcolorbox}
\usepackage{amsmath}
\usepackage{etoolbox}
\usepackage{url}
\usepackage{tabularx}
\usepackage{fancyhdr}

\usepackage{array}
\AtBeginEnvironment{verbatim}{\scriptsize}
\usepackage{hyperref}
\hypersetup{
    colorlinks=true,
    linkcolor=blue,
    filecolor=magenta,
    citecolor=cyan,
    urlcolor=cyan,
    breaklinks=true,
    }

\def\BibTeX{{\rm B\kern-.05em{\sc i\kern-.025em b}\kern-.08em
    T\kern-.1667em\lower.7ex\hbox{E}\kern-.125emX}}

\newlength{\tightskip}
\let\oldnormalsize\normalsize
\renewcommand{\normalsize}{%
    \oldnormalsize
    \setlength{\abovedisplayskip}{\tightskip}%
    \setlength{\belowdisplayskip}{\tightskip}%
    \setlength{\abovedisplayshortskip}{\tightskip}%
    \setlength{\belowdisplayshortskip}{\tightskip}%
}

\fancypagestyle{firstpage}{
  \fancyhf{} 
  \fancyfoot[C]{\footnotesize
    \begin{minipage}{\textwidth}
    \centering
    \color{black}{
© 2026 IEEE. This paper has been accepted for presentation at the 2026 IEEE International Conference on Communications,
Control, and Computing Technologies for Smart Grids (SmartGridComm)}
    \end{minipage}
  }

}

\begin{document}

\title{Explainable Post-Disaster Grid Observability Recovery Using Human-Oversight Agentic LLMs\\
}

\author{
\IEEEauthorblockN{
Biswas Rudra Jyoti Arka\textsuperscript{1},
Sadman Sakib\textsuperscript{2},
Md. Zahidul Islam\textsuperscript{1,*},
Shamsun Nahar Edib\textsuperscript{2}
}
\IEEEauthorblockA{\textsuperscript{1}School of Electrical, Computer, and Biomedical Engineering, Southern Illinois University Carbondale, IL 62901, USA}
\IEEEauthorblockA{\textsuperscript{2}Department of Electrical and Computer Engineering, Montana State University, Bozeman, MT 59717, USA}
\IEEEauthorblockA{\textsuperscript{*}Corresponding author: mdzahidul.islam@siu.edu}
}

\maketitle

\thispagestyle{firstpage}

\begin{abstract}

Post-disaster phasor measurement unit (PMU) outages reduce power-system observability and degrade operator situational awareness, requiring sequential restoration under limited resources. Existing PMU restoration methods based on optimization or heuristics can generate restoration schedules, but they often provide limited support for explanation, traceability, and operator interaction. This paper proposes an agentic tool-calling framework orchestrated by a large language model (LLM) for post-disaster PMU restoration and grid observability recovery. In this framework, the LLM does not directly solve the restoration optimization problem; instead, it coordinates validated backend tools required for post-disaster restoration, including observability assessment, restoration planning, state updates, and operator verification. The framework also maintains a structured tool-call history and execution context that keep restoration decisions traceable and explainable, while enabling context-aware operator question answering during the restoration process. Simulation results on IEEE 30-bus and IEEE 57-bus systems show that the proposed framework achieves observability recovery comparable to a mixed-integer linear programming (MILP) solution, while providing tool-grounded explanations, interactive operator support, and human-overseen execution.
\end{abstract}


\vspace{-8pt}

\section{Introduction}

Modern power-system operation relies on timely and accurate measurements to maintain situational awareness, support state estimation, and guide corrective actions. Phasor Measurement Units~(PMUs) are an important part of this measurement infrastructure because they provide GPS-synchronized voltage and current phasors for wide-area monitoring, protection, and control~\cite{phadke2018pmu}. When the available PMU and sensor measurements are sufficient to observe all buses in the network, the system can support reliable state estimation and downstream operational decisions~\cite{baldwin1993pmu}.

During extreme events, this measurement capability can be severely degraded. Natural disasters, such as hurricanes and wildfires, or coordinated cyber-physical attacks on measurement infrastructure, may disable multiple PMUs across the grid~\cite{liang2017fdi}. The resulting loss of observability can leave operators with limited visibility of the system at the exact time when restoration and corrective decisions are most critical. However, failed PMUs cannot usually be restored all at once because field crews, spare devices, and communication resources are limited during the post-event period. Therefore, PMU restoration becomes a sequential decision-making problem, where the restoration order determines how quickly observability is recovered under resource constraints~\cite{lei2019resilient,edib2023cyber}.

Existing PMU restoration approaches commonly rely on mixed-integer linear programming~(MILP) formulations~\cite{edib2021optimal}, greedy methods, or heuristics~\cite{haggi2020multi}. These methods are effective for computing restoration sequences when the objective function, constraints, and system conditions are clearly defined. However, they are usually implemented as standalone optimization modules that return restoration decisions with limited support for explanation and interaction. In practical post-disaster operation, operators may need to examine why a specific PMU or zone is prioritized and incorporate their own expertise. Therefore, beyond computing a restoration sequence, an operator-facing workflow is needed to make the restoration process traceable, explainable, and interactive.

Recent advances in large language models~(LLMs) have shown strong capability in language understanding, reasoning over contextual information, and generating human-readable explanations. However, direct use of LLMs for critical power-system decision making is limited by their non-deterministic behavior and tendency to generate unsupported numerical or procedural outputs when complex calculations are required~\cite{reason_fail}. Agentic AI addresses part of this limitation by enabling LLMs to follow task-specific instructions, call external tools, retrieve structured outputs, and coordinate multi-step workflows~\cite{agenticAI,agenticAI2}. This makes agentic LLMs suitable for operator-facing restoration tasks, where the reasoning and explanation capability of the LLM can be combined with validated external computation tools to support traceable and human-supervised decision making. Recent studies have explored agentic AI for power-system analysis tasks, such as AC optimal power flow, contingency analysis, and interconnection studies \cite{jin2025gridmind, zhang2025grid}. However, existing approaches have not considered agentic AI for sensor restoration problems.

To overcome the above limitations, this paper proposes an agentic tool-calling LLM framework for post-disaster PMU restoration and system observability recovery. The proposed method treats the LLM as a workflow orchestrator and explanation interface, while restoration computations are performed by validated external tools used in power-system restoration problems. Specifically, the framework decomposes the PMU restoration process into modular callable operations, including a rolling-horizon MILP function, coordinates them through a guarded tool-calling workflow, and maintains a trace of intermediate outputs for operator inspection. At each restoration step, the LLM explains the tool-generated restoration plan and supports human oversight before the action is applied. In addition, the framework supports context-aware interactive question answering, allowing operators to query the current restoration state, priority zones, and remaining observability gaps during the restoration process. This capability is not naturally provided by standalone optimization modules and provides an operator-facing decision-support layer for post-disaster restoration.

\section{Problem Statement}

\label{sec:problem_statement}

Post-disaster PMU restoration is a sequential decision-making problem in which failed PMUs must be restored over multiple steps to recover grid observability under limited restoration resources. At each step, the restoration process must determine which PMUs should be restored while preserving already active PMUs, respecting the available resource budget, and improving system observability.

The inputs to the problem are the installed PMU locations, post-disaster failed PMUs, known post-disaster power-grid topology, and available restoration resources, such as repair crews or spare devices. The output is a PMU restoration sequence that maximizes observability improvement over the restoration process, with only a limited number of PMUs restored at each step according to the available resources.


This paper focuses on the post-disaster restoration of failed PMUs, where PMU locations are predefined, e.g., obtained by solving a PMU placement problem.




\section{Proposed Framework}
\label{sec:framework}

\begin{figure}
    \centering
    \includegraphics[width=.9\linewidth]{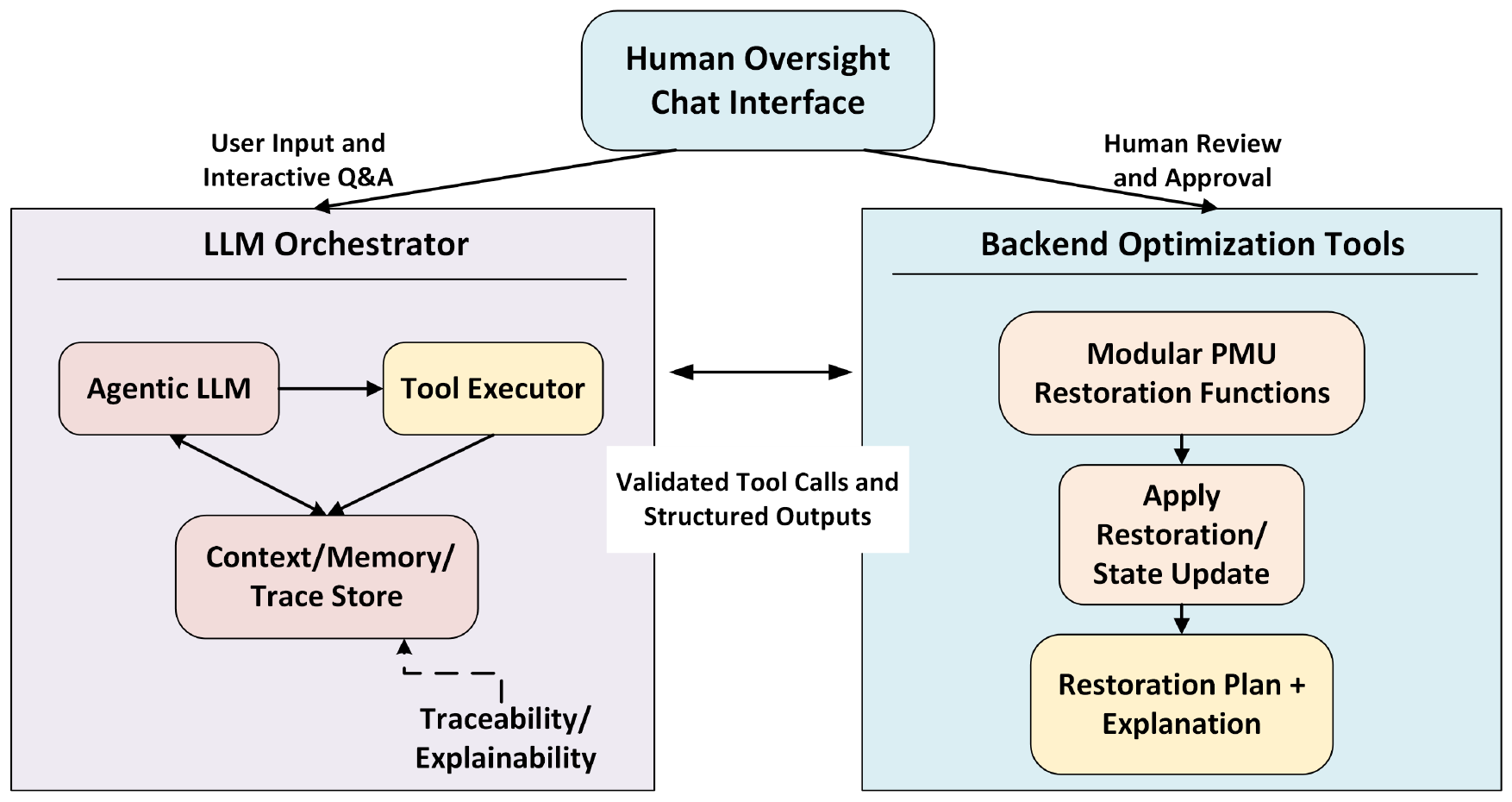}
    \caption{Overview of the proposed PMU restoration framework.}
    \label{fig:Algorithm}
\end{figure}

An overview of the proposed LLM-based agentic AI workflow is shown in Fig.~\ref{fig:Algorithm}. The workflow integrates a MILP-based PMU restoration solution with an agentic LLM, combining the strengths of classical optimization and LLM-based orchestration. The workflow begins with user-defined instructions, which are maintained in the framework context. Based on the instruction and task, the LLM extracts the required information and invokes the backend optimization tools through the executor to obtain a post-disaster restoration plan. The LLM also provides an explanation accompanying the restoration plan. The human operator interacts with the LLM to review, approve, or modify the plan. Upon receiving approval, the system state is updated based on the approved plan, assuming that the corresponding restoration action is implemented in the field by repair crews. The framework maintains a well-structured context memory that provides traceability, interactive Q\&A, and operator-facing explainability. In the following subsections, we first define the backend tool-callable restoration functions and then describe the LLM orchestration framework.

\subsection{Tool-Callable PMU Restoration Functions}

\label{sec:functions}

Through a systematic modular function design, the proposed framework separates power-system restoration computation from LLM-based orchestration. Core restoration operations are implemented as deterministic tool-callable functions with structured inputs and outputs. The modular functions are summarized in Table~\ref{tab:functions}. Each function performs a restoration-specific task and produces outputs required by subsequent functions in the restoration pipeline.

\textbf{Zone and PMU representation:}
For a network partitioned into $Z$ zones, zone $r$ is represented by its bus set $\mathcal{R}_r$, PMU-candidate set $\mathcal{P}_r \subseteq \mathcal{R}_r$, PMU status vector $\mathbf{s}_r \in \{0,1\}^{|\mathcal{P}_r|}$, and zonal graph $G_r=(\mathcal{R}_r,\mathcal{E}_r)$. The set of inter-zone tie-lines is denoted by $\mathcal{T}$.

PMU placement problems have been extensively studied in the existing literature and can be used to obtain the initial PMU locations required for observability. 
For PMU-based observability, a bus $i$ in zone $r$ is observable if the bus itself or any of its neighboring buses is equipped with a PMU~\cite{edib2021optimal}. Following this observability condition, a connectivity matrix
$C_r \in \{0,1\}^{|\mathcal{R}_r|\times|\mathcal{P}_r|}$ is constructed for each zone and used in the MILP-based restoration function. An entry of 1 in $C_r$ indicates that the corresponding PMU candidate can observe the corresponding bus; otherwise, the entry is 0.

\begin{table*}[t]
\caption{Summary of modular tool-callable PMU restoration functions.}
\label{tab:functions}
\centering
\footnotesize
\renewcommand{\arraystretch}{1.2}
\begin{tabular}{@{}p{0.28\linewidth}p{0.3\linewidth}p{0.15\linewidth}p{0.2\linewidth}@{}}
\toprule
\textbf{Function} & \textbf{Role} & \textbf{Input} & \textbf{Output} \\
\midrule
\texttt{partition\_network}
& Zone partitioning, PDC-site identification, and tie-line extraction
& Case id, $Z$
& $\{\mathcal{R}_r,\mathcal{P}_r,G_r\}$, $\mathcal{T}$ \\

\texttt{zone\_weights}
& Assign zone criticality weights
& Zones $Z$
& $\{W_r\}$ \\

\texttt{generate\_failure\_scenario}
& Simulate post-disaster PMU outages for experiments
& $\{\mathcal{P}_r\}$, failure rate
& $\{\mathbf{s}_r\}$ \\

\texttt{compute\_zone\_scores\_and\_budgets}
& Compute zone scores and allocate restoration budget
& $\{W_r,O_r\}$, total budget $B$, $\alpha$, $\beta$
& $\{I_r\}$, $\{b_r\}$ \\

\texttt{propose\_restoration}
& Solve rolling-horizon MILP and propose restoration plan
& $\{\mathcal{P}_r,\mathbf{s}_r,C_r\}$, $\{b_r\}$, $H$
& Proposed PMU updates \\

\texttt{apply\_restoration}
& Commit operator-approved restoration updates
& Proposed PMU updates
& Updated $\{\mathbf{s}_r\}$ \\

\texttt{check\_total\_observability}
& Compute zone-level and grid-wide observability
& $\{C_r,\mathbf{s}_r\}$
& $\{O_r\}$, total observability status \\

\texttt{evaluate\_tieline\_observability}
& Check inter-zone tie-line PMU coverage
& $\mathcal{T}$, $\{\mathbf{s}_r\}$
& Tie-line observability status \\
\bottomrule
\end{tabular}
\end{table*}

\textbf{Rolling-horizon MILP:}
The main restoration decision is computed by \texttt{propose\_restoration}, which solves a rolling-horizon MILP for each zone. In the rolling-horizon MILP, a horizon of $H$ future steps is considered instead of solving the full restoration problem over all remaining steps, while only the first-step decision is proposed for operator inspection. This provides a faster restoration solution while maintaining high restoration quality. The horizon then advances iteratively, using the updated PMU state as the initial state, and the problem is re-solved. Let $\mathbf{x}_t\in\{0,1\}^{|\mathcal{P}_r|}$ denote the PMU status vector and $\mathbf{y}_t\in\{0,1\}^{|\mathcal{R}_r|}$ denote the bus observability vector at step $t$. Let $b_r$ denote the restoration budget, such as the number of available crews, allocated to zone $r$. The rolling-horizon PMU restoration problem can be formulated as:
\begin{subequations}
\label{eq:milp}
\begin{align}
  \max_{\mathbf{x}_t,\mathbf{y}_t} \quad
  & \sum_{t=1}^{H} \mathbf{1}^{\top}\mathbf{y}_t
  \label{eq:milp_obj}\\
  \text{s.t.} \quad
  & \mathbf{y}_t \le C_r \mathbf{x}_t,
  && \forall t
  \label{eq:milp_obs}\\
  & \mathbf{x}_1 \ge \mathbf{s}_r,
  &&
  \label{eq:milp_active}\\
  & \mathbf{1}^{\top}(\mathbf{x}_1-\mathbf{s}_r)
  \le b_r,
  &&
  \label{eq:milp_budget_first}\\
  & \mathbf{1}^{\top}(\mathbf{x}_t-\mathbf{x}_{t-1})
  \le b_r,
  && \forall t>1
  \label{eq:milp_budget_future}\\
  & \mathbf{x}_t \ge \mathbf{x}_{t-1},
  && \forall t>1.
  \label{eq:milp_monotone}
\end{align}
\end{subequations}
Constraint~\eqref{eq:milp_obs} ensures the PMU-based observability condition.
Constraint~\eqref{eq:milp_active} preserves the currently active PMUs,
while \eqref{eq:milp_budget_first} limits the number of newly restored
PMUs in the first committed step. Constraint~\eqref{eq:milp_budget_future}
enforces the restoration budget over later horizon steps, and
\eqref{eq:milp_monotone} prevents an active PMU from being deactivated.
The function \texttt{propose\_restoration} returns the MILP-suggested
restoration plan, and \texttt{apply\_restoration} commits the
operator-approved plan to the system state. 

The above formulation provides a near-optimal PMU restoration plan at each restoration step while satisfying the restoration constraints. The agentic LLM calls this trusted function as a tool, preserving power-system-grounded decision making and reducing unsupported LLM outputs.

\textbf{Zone importance and budget allocation:}
The function \texttt{zone\_weights} assigns an importance weight $W_r$ to each zone to emulate heterogeneous criticality, such as zones containing hospitals or other critical infrastructure. The function \texttt{compute\_zone\_scores\_and\_budgets} combines this weight with the current zone observability to allocate limited restoration resources. The observability of zone $r$ is:
\begin{equation}
O_r = \frac{1}{|\mathcal{R}_r|}
\sum_{i\in\mathcal{R}_r} o_i,
\label{eq:zone_obs}
\end{equation}
and its dynamic priority score is
\begin{equation}
I_r = \alpha W_r + \beta(1-O_r),
\qquad \alpha+\beta=1.
\label{eq:zone_score}
\end{equation}
Based on the zone priority scores, this function also allocates the restoration budget such that zones with low observability and high importance receive higher budgets, while a fully observable zone receives no budget. This guides the LLM to determine the zones to prioritize during the restoration process.

\textbf{Additional functions:}
The function \texttt{check\_total\_observability} reports zonal and system-wide observability, which guides the LLM orchestrator in deciding whether to stop or continue the restoration process. The function \texttt{evaluate\_tieline\_observability} checks whether tie-lines among zones are observable.

The function \texttt{generate\_failure\_scenario} can be called by the LLM in simulation to generate random failure scenarios and emulate post-disaster PMU outages; in real operation, this would be replaced by actual PMU availability data. The function \texttt{partition\_network} divides the network into multiple zones using $k$-medoids clustering over the shortest-path distance matrix.

\subsection{Agentic Tool-Calling Framework for PMU Restoration}

\label{sec:llm}

Building on the tool-callable functions in Section~\ref{sec:functions}, this subsection describes how the LLM interacts with the deterministic restoration tools through a structured tool-calling interface, an executor-mediated context, and human oversight. 
\subsubsection{Tool-Calling Interface and Executor Context}

The modular functions in Table~\ref{tab:functions} are exposed to the LLM through a structured tool-calling interface. Each tool declaration specifies the function name, natural-language description, JavaScript Object Notation (JSON) input-parameter schema, and required arguments. The framework also includes an \emph{executor}, which serves as an interface layer between the LLM and the deterministic backend functions. Through the structured tools and the executor, all numerical computation, MILP solving, observability checking, and state updates are performed by deterministic backend functions rather than directly by the LLM.
A tool declaration is represented as:
\begin{equation}
\tau_i = \{ n_i,\, d_i,\, X_i,\, C_i \},
\label{eq:tool_definition}
\end{equation}
where $n_i$ is the tool name, $d_i$ is the tool description, $X_i$ is the JSON input-parameter schema, and $C_i$ denotes the required arguments and execution constraints. Given an LLM-provided argument object $x_i$ and the executor-side context $c_t$, a valid tool call executes:
\begin{equation}
y_i = f_i(x_i,c_t),
\label{eq:tool_execution}
\end{equation}
where $f_i(\cdot)$ is the backend function and $y_i$ is the tool-specific JSON output.

During each tool call, the executor receives the tool name and JSON arguments from the LLM, validates the call, fills missing fields from the stored context when available, executes the corresponding backend function, and returns the JSON result. Let $c_t$ denote the executor-side context and $m_t$ denote the LLM-visible message history at step $t$. The returned output is appended to the LLM-visible message history and is also used to update the executor-side context:
\begin{equation}
c_{t+1}=U(c_t,y_i), \qquad
m_{t+1}=m_t \oplus \{y_i\},
\label{eq:context_update}
\end{equation}
where $U(\cdot)$ denotes the executor-side context update and $\oplus$ denotes appending to the ordered message history. Since different functions perform different restoration tasks, their returned fields are function-specific. For example, \texttt{check\_total\_observability} returns zone-level and grid-wide observability status, whereas \texttt{propose\_restoration} returns the MILP-based restoration plan.

The framework maintains two forms of context. The LLM-visible message history $m_t$ stores the system instruction, user request, tool-call requests, and returned tool outputs. This enables the LLM to follow the user-defined workflow, select the appropriate tool-callable function during the restoration process, and generate the restoration report. In parallel, the executor maintains a structured context dictionary $c_t$ containing specified run parameters and restoration states, such as the restoration budget, failure rates, zone weights, zone networks, PMU candidate buses, PMU ON/OFF states, zone scores, allocated budgets, restoration plans, and observability status.

The operator remains in the loop by reviewing the proposed restoration plan before implementation and can approve or modify the plan. The execution history records tool names, inputs, outputs, explanations, restoration updates, and observability changes, forming a traceable log of the rolling-horizon restoration process.

\subsubsection{Guarded Tool Calling with Human Approval}

The LLM is used as a controlled interface between the operator and the restoration tools, with the main components summarized in Table~\ref{tab:agent_workflow}. The system instruction defines the agent's role, execution rules, tool-call order, stopping condition, grounding requirement, and final-report format. The user task prompt initiates the specific restoration run and requests the final restored PMUs and observability status. During execution, the LLM is instructed to use the specified run arguments, prior tool outputs, and stored context.

\begin{table}[t]
\centering
\caption{Components and safeguards in guarded tool calling.}
\label{tab:agent_workflow}
\footnotesize
\setlength{\tabcolsep}{3pt}
\renewcommand{\arraystretch}{1.12}
\begin{tabular}{p{0.30\linewidth}p{0.60\linewidth}}
\toprule
\textbf{Component} & \textbf{Function} \\
\midrule
System instruction
& Defines the LLM role, tool-call order, stopping condition, grounding rule, rationale requirement, and final-report format. \\

User task prompt
& Initiates the restoration run and requests the restored PMUs, restoration order, and final observability status. \\

LLM agent
& Converts operator requests into admissible tool calls, explains tool outputs, and answers follow-up questions using stored context. \\

Executor
& Validates tool names and arguments, fills required inputs from context, executes backend functions, updates context, and records the trace. \\

Backend tools
& Compute zone scores, restoration budgets, MILP-based restoration plans, state updates, and observability checks. \\

Context memory
& Stores tool inputs, outputs, PMU states, zone scores, budgets, applied actions, observability status, and operator feedback. \\

Human operator
& Reviews each proposed restoration action before application and may approve or modify the plan. \\
\bottomrule
\end{tabular}
\end{table}

\begin{table}
\centering
\caption{Representative LLM tool-call trace for restoration proposal.}
\label{tab:tool_call_trace}
\footnotesize
\setlength{\tabcolsep}{3pt}
\renewcommand{\arraystretch}{1.05}
\begin{tabular}{p{0.18\columnwidth} p{0.74\columnwidth}}
\hline
\textbf{Item} & \textbf{Content} \\
\hline
Tool call 
& \texttt{propose\_restoration} \\

Input 
& \texttt{\{"n\_zones":4, "budgets":[1,0,1,1]\}} \\

Tool output 
& Zone 0: activate PMU [50], observable buses = 19; 
Zone 1: activate none, observable buses = 7; 
Zone 2: activate PMU [51], observable buses = 19; 
Zone 3: activate PMU [2], observable buses = 12. \\

LLM rationale 
& Continued budget allocation to Zones 0 and 2 is needed to improve observability, while Zone 1 requires no additional budget. Zone 3 receives one PMU to increase coverage. \\
\hline
\end{tabular}
\end{table}
These components work together as follows. The system instruction constrains the LLM to request only admissible restoration tools, while the user task prompt starts the specific restoration case. In each restoration round, the LLM requests the required scoring, restoration-planning, or verification tool using the current context. The executor validates the tool name and arguments, executes the corresponding backend function, records the output, and updates the context memory.
The LLM then explains the returned restoration result using tool-generated zone scores, allocated budgets, selected PMUs, and observability status. Before the PMU state is updated, the human operator can review or modify the restoration plan. 

After the state update, observability is re-evaluated, the updated state is stored in context memory, and the process continues until full observability is achieved.

\subsubsection{Context-Based Explanation and Interaction}

The framework provides explainability at multiple stages. After each restoration tool call, the LLM generates a natural-language rationale using the stored context, including zone scores, allocated budgets, selected PMUs, and updated observability status. The explanation is therefore grounded in tool outputs and produced together with the restoration step, rather than added as an independent post-hoc interpretation. At the end of the workflow, the LLM also summarizes the restored PMUs, their restoration order, the final observability status, and any unresolved observability gaps. Because the execution history records tool inputs, tool outputs, explanations, restoration actions, and observability changes, each explanation can be traced back to the corresponding tool call and decision step.

The stored context also enables operator interaction. During or after restoration, the operator can ask follow-up questions about the current observability state, priority zones, remaining gaps, or previous restoration decisions. Thus, the framework provides context-aware explanation and operator support instead of only returning a one-shot optimization result.

\begin{table*}[t]
\centering
\caption{Context-aware operator interaction during restoration.}
\label{tab:interactive_qa}
\footnotesize
\setlength{\tabcolsep}{3pt}
\renewcommand{\arraystretch}{1.08}
\begin{tabularx}{\textwidth}{
>{\raggedright\arraybackslash}p{0.23\textwidth}
>{\raggedright\arraybackslash}X
>{\raggedright\arraybackslash}p{0.16\textwidth}}
\hline
\textbf{Operator query} & \textbf{Agent response summary} & \textbf{Capability} \\
\hline
Which zone has the lowest observability?
& Identifies Zone 2 as the least observable zone because it has only one active PMU, at index 4.
& State retrieval \\

Which zone should have the highest priority?
& Identifies Zone 0 as the highest-priority zone due to its high weight of 10.0 and reported score of 0.3143.
& Priority reasoning \\

Is the grid fully observable now?
& Reports that the grid is not fully observable and that observability gaps remain in the current restoration state.
& Situation assessment \\

Which zone still needs more restoration?
& Identifies Zone 0 as requiring more restoration based on its reported score of 0.7714, high weight, and previous budget allocation.
& Context tracking \\

Are all zones observable now?
& Reports that Zone 1 is fully observable, while Zone 0 and Zone 2 still have residual gaps with scores of approximately 0.886 and 0.923, respectively.
& Multi-turn monitoring \\
\hline
\end{tabularx}
\end{table*}

\begin{table*}[t]
\centering
\caption{Expert consistency check of LLM-generated rationales.}
\label{tab:rationale_eval}
\footnotesize
\setlength{\tabcolsep}{3pt}
\renewcommand{\arraystretch}{1.08}
\begin{tabularx}{\textwidth}{
>{\raggedright\arraybackslash}p{0.15\textwidth}
>{\centering\arraybackslash}p{0.04\textwidth}
>{\raggedright\arraybackslash}p{0.31\textwidth}
>{\raggedright\arraybackslash}X
>{\centering\arraybackslash}p{0.08\textwidth}}
\hline
\textbf{Case} & \textbf{Iter.} & \textbf{Human reference rationale cue} & \textbf{LLM rationale cue} & \textbf{Check} \\
\hline
IEEE 57-bus, 3 zones, budget 4, 90\% failure
& 1
& Zone 0 has the highest priority score and demands more budget than the other zones.
& Allocates budgets $[2,1,1]$ and explains that Zone 0 has the highest importance score and receives the largest budget.
& Consistent \\

& 2
& Zone 0 has many unobservable buses and requires the highest budget due to its priority score.
& Allocates budgets $[2,1,1]$ and explains that Zone 0 has the highest score and therefore receives the highest budget.
& Consistent \\

& 3
& All zones have close observability and importance values, but Zone 0 has the highest computed score and receives higher budget.
& Allocates budget 2 to Zone 0 and budget 1 to Zones 1 and 2, reflecting Zone 0's higher score.
& Consistent \\

& 4
& Zone 1 has reached full observability, so the budget is divided between Zones 0 and 2.
& Allocates budgets to Zones 0 and 2 while assigning no budget to Zone 1 because Zone 1 is fully observable.
& Consistent \\

& 5
& Only Zone 0 remains not fully observable, so the whole budget is allocated to Zone 0.
& Allocates budget 4 to Zone 0 and no budget to Zones 1 and 2 because they are already fully observable.
& Consistent \\
\hline
IEEE 30-bus, 4 zones, budget 3, 90\% failure
& 1
& Zone 0 has the lowest score, so the available budget is distributed to the other three zones.
& Restores one PMU in each of Zones 1, 2, and 3 to improve observability in the least observable zones.
& Consistent \\

& 2
& Zones 1 and 3 are fully observable; budget is assigned to Zones 0 and 2, with more emphasis on Zone 2.
& Explains that restoration should focus on Zones 0 and 2 because they still need improvement.
& Consistent \\

& 3
& Only Zone 0 remains not fully observable, so the entire budget is allocated to Zone 0.
& Explains that Zone 0 remains partially observable and the other zones are already fully observable.
& Consistent \\
\hline
\multicolumn{2}{l}{\textbf{Total}} 
& \multicolumn{2}{l}{8 evaluated iterations; no unsupported PMU indices, bus numbers, or budget values were observed.}
& 8/8 \\
\hline
\end{tabularx}
\end{table*}

\vspace{-2pt}
\section{Case Study} 
\label{sec:casestudy}
The proposed framework is validated on the IEEE 30-bus and IEEE 57-bus systems under multiple scenarios with varying failure rates, restoration budget $b_k$, and number of zones. Failure rates of 30\%, 50\%, 70\%, and 90\% are considered. The restoration budget is varied from 1 to 4, and the number of partitioned zones is varied from 1 to 4. For the MILP optimization, the prediction horizon is set to $H=3$. Although other LLMs can be adopted, the proposed framework is tested using OpenAI's GPT-4o mini model, which is a fast and cost-effective model with reasoning capability~\cite{openai_gpt4}.


\subsection{Agent-Layer Capability and Value Demonstration}
In this subsection, we evaluate the capabilities and added value introduced by the agentic LLM layer beyond the deterministic MILP-based restoration tool.

\subsubsection{LLM tool-call and explanation trace}
To illustrate how the agent interacts with deterministic restoration tools, 
Table~\ref{tab:tool_call_trace} presents a representative tool-call trace from the IEEE 57-bus case. The LLM invokes the MILP-based \texttt{propose\_restoration} tool using the current zone-budget vector. The tool returns the restoration action for the current step, and the LLM generates a rationale grounded in the returned output for operator review. 
This trace illustrates the separation between deterministic computation and language-based explanation. The PMU restoration action is produced by the MILP tool, while the LLM explains the returned plan for operator review. This design reduces the risk of unsupported restoration decisions being generated directly by the LLM. The complete tool-call traces are provided in the project repository~\cite{github_repo}.

\subsubsection{Context-aware operator interaction}
To demonstrate interactive operator support, we evaluate whether the agent can answer follow-up questions during the restoration process using the stored execution context. Table~\ref{tab:interactive_qa} summarizes one representative multi-turn interaction. The questions are asked during the restoration process, and the answers are generated from the available context, including active PMU states, zone observability, zone weights, previous restoration actions, and remaining observability gaps.
Table~\ref{tab:interactive_qa} shows that the agent can answer multi-turn operator questions using the evolving restoration context. This demonstrates that the proposed framework is not limited to producing a final static report after a fixed workflow. Instead, the operator can inspect the current system state, restoration priority, and remaining observability gaps through natural-language follow-up questions.

\subsubsection{Expert rationale consistency evaluation}
We further evaluate whether the LLM-generated rationales are consistent with human-written reference rationales. This evaluation is intended as a faithfulness check rather than a statistical user study. For each restoration iteration, a human reference rationale was written using the tool-generated budget allocation, zone scores, and observability status. The LLM rationale was then compared against this reference. A rationale is considered consistent if it identifies the same priority zone or budget-allocation logic and does not introduce unsupported PMU indices, bus numbers, or budget values.

Table~\ref{tab:rationale_eval} reports the key rationale cues from both the human reference and the LLM output for each evaluated iteration. Two representative scenarios are considered: IEEE 57-bus with 3 zones, budget 4, and 90\% PMU failure rate; and IEEE 30-bus with 4 zones, budget 3, and 90\% PMU failure rate.
Table~\ref{tab:rationale_eval} shows that the LLM rationales follow the same budget-allocation logic as the human reference rationales across all evaluated iterations. 
This result shows that the agent layer translates deterministic restoration outputs into explanations that remain consistent with the underlying tool results and human reference reasoning, while avoiding unsupported restoration decisions.





\subsection{Restoration Performance Evaluation}
This subsection evaluates the framework's PMU restoration and observability recovery performance under different post-disaster failure scenarios for the IEEE 30- and 57-bus systems. The user prompt defines the required tool-call sequence and instructs the framework to continue restoration until full observability is achieved or a maximum iteration limit is reached. The prompt is provided in the project repository~\cite{github_repo}.

\begin{figure}[t]
    \centering
    \begin{subfigure}{.75\linewidth}
        \centering
        \includegraphics[width=\linewidth]{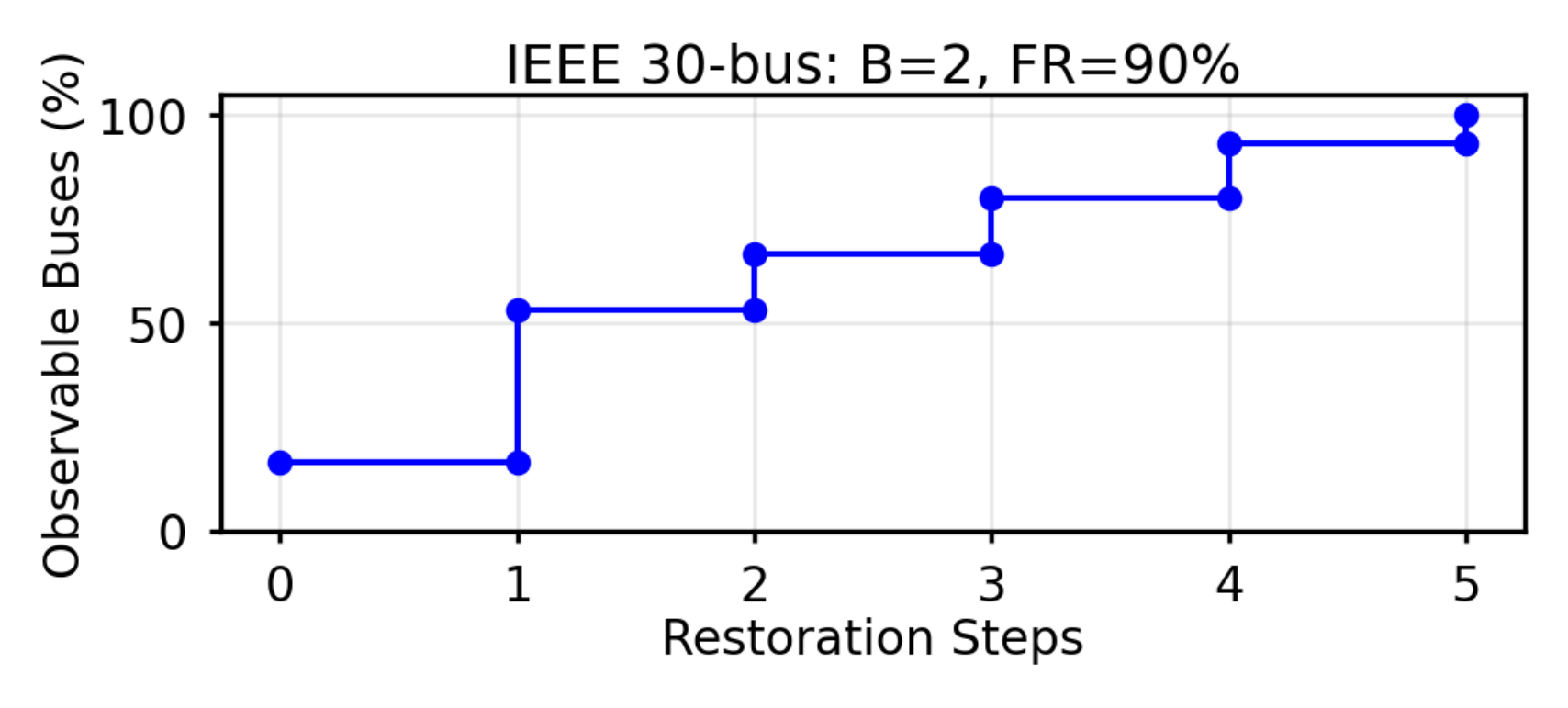}
        \label{fig:case30}
    \end{subfigure}

    \vspace{-18pt}
    \begin{subfigure}{.75\linewidth}
        \centering
        \includegraphics[width=\linewidth]{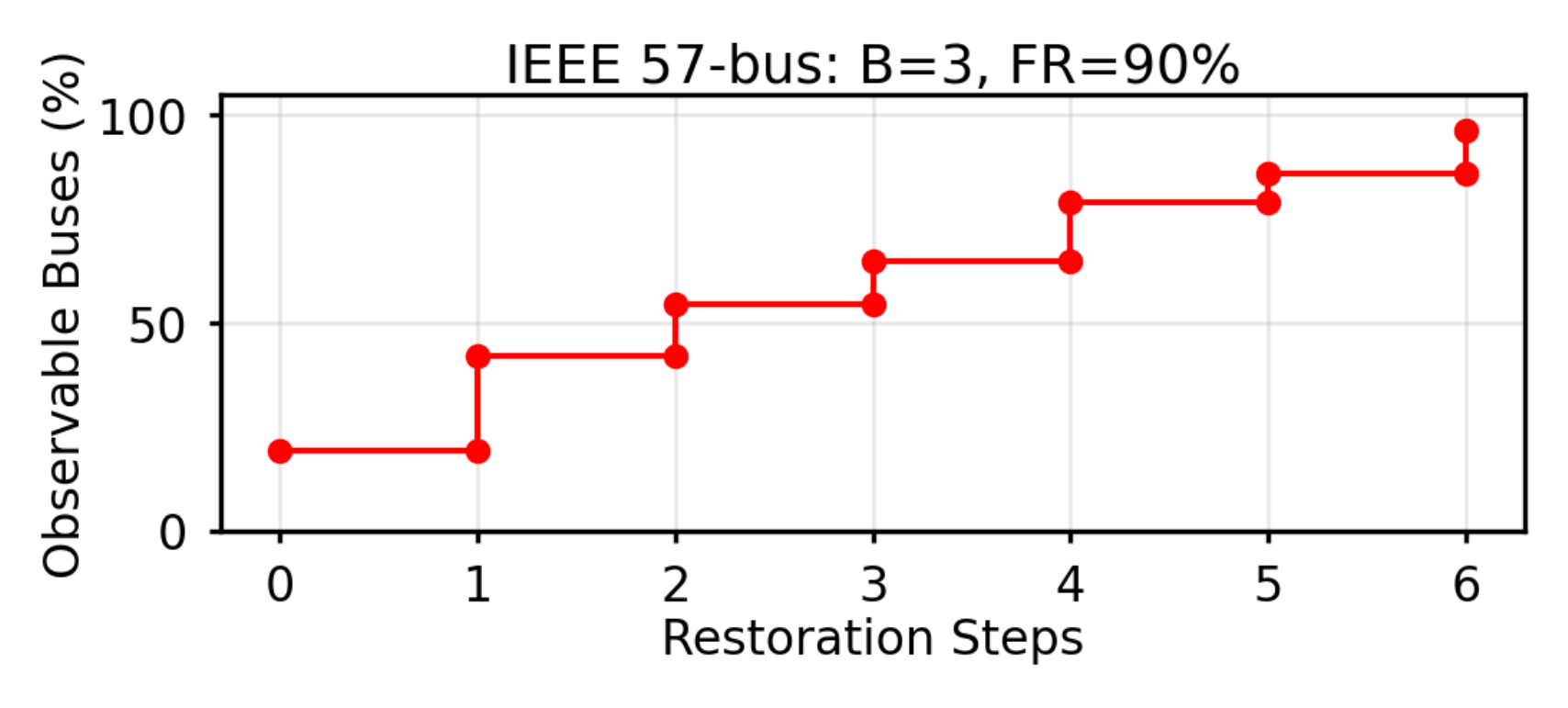}
        \label{fig:case57}
    \end{subfigure}
    \vspace{-15pt}
    \caption{Observability vs. restoration performance. }
    \label{fig:obs_vs_res}
\end{figure}

Fig.~\ref{fig:obs_vs_res} shows the restoration trajectory for the IEEE 30-bus and IEEE 57-bus systems with 4 zones, budget 3, and 90\% failure rate. In both cases, the overall observability increases gradually after each restoration step. This indicates that the framework continues to call the backend optimization tools, updates the restoration state after each approved action, and iteratively improves observability until full observability is achieved, as instructed by the user prompt.

\begin{figure}[t]
    \centering
    \includegraphics[width=.85\linewidth]{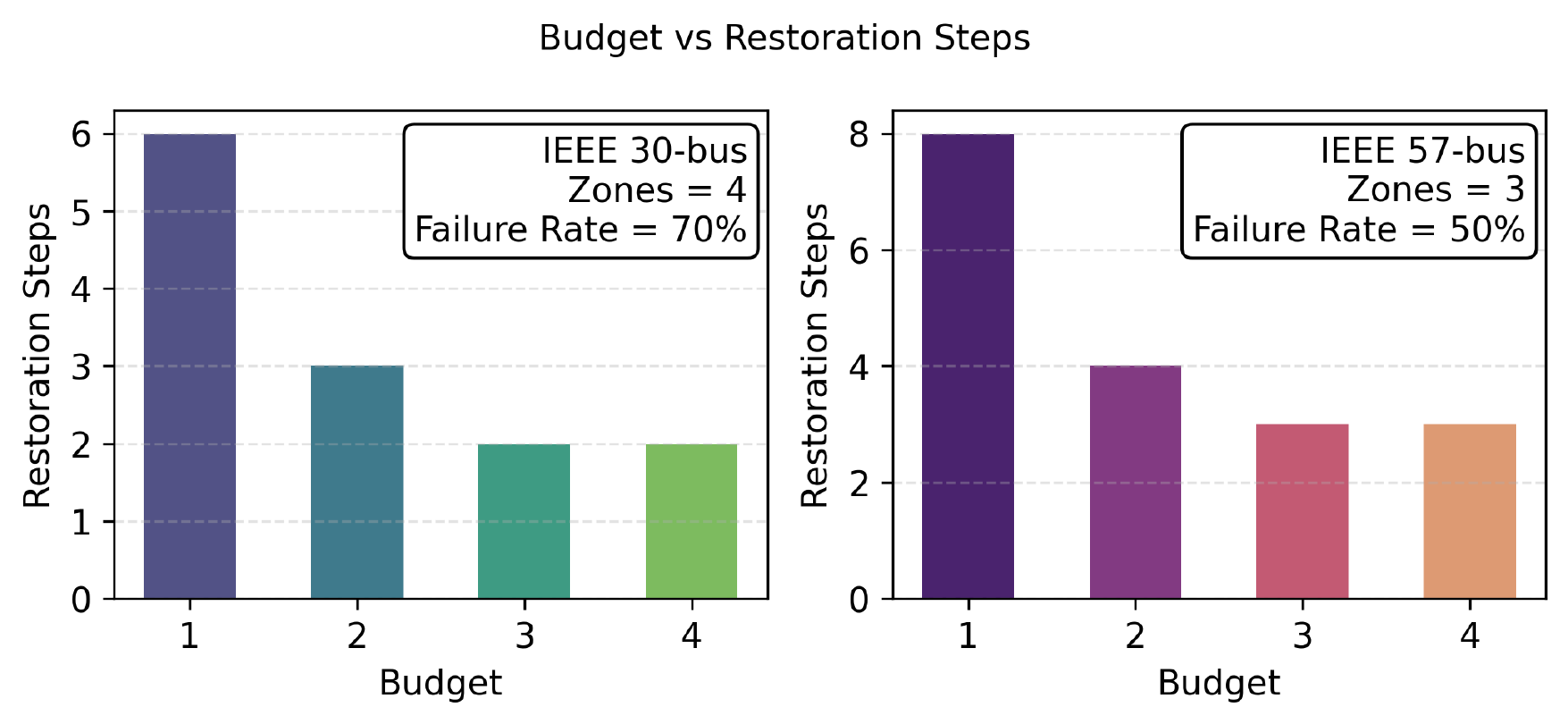}
    \caption{Restoration performance under different budgets.}
    \label{fig: budget vs res}
\end{figure}

Fig.~\ref{fig: budget vs res} shows the effect of restoration budget on restoration performance. As the available budget increases, fewer restoration steps are required to recover observability. This result confirms that the framework can handle different budget scenarios and allocate the available restoration resources across zones based on their priority. Therefore, when more resources are available, the framework can restore more PMUs in each step and accelerate observability recovery.

\begin{figure}[t]
    \centering
    \includegraphics[width=.9\linewidth]{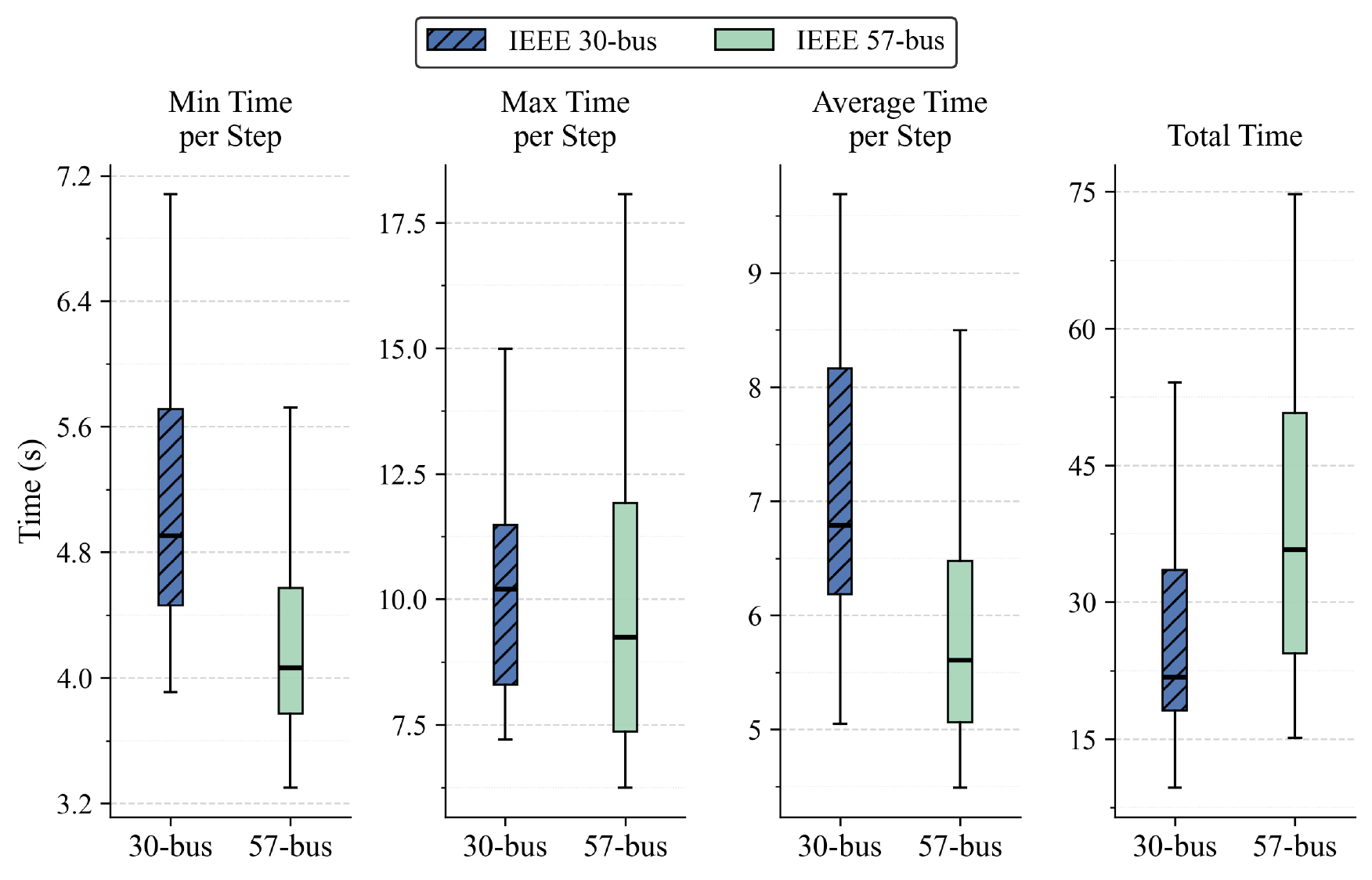}
\caption{Runtime of the restoration plan, including LLM orchestration and backend optimization calls, for IEEE 30-bus and IEEE 57-bus systems. From left to right, the boxplots show the minimum, maximum, average per-step, and total runtime.}
    \label{fig:time_plot}
\end{figure}

Fig.~\ref{fig:time_plot} reports the minimum, maximum, and average runtime per restoration step, along with the total workflow time required to achieve full observability. The reported times include LLM-based tool orchestration and backend optimization on a personal computer with an Intel Core Ultra 7 processor and 64 GB of RAM, but exclude field-crew execution time. The results show that, despite the additional LLM orchestration and explanation steps, the framework maintains decision times of only a few seconds per restoration step in the tested cases, which is practical for post-disaster restoration planning.

\begin{figure}[t]
    \centering
    \includegraphics[width=.85\linewidth]{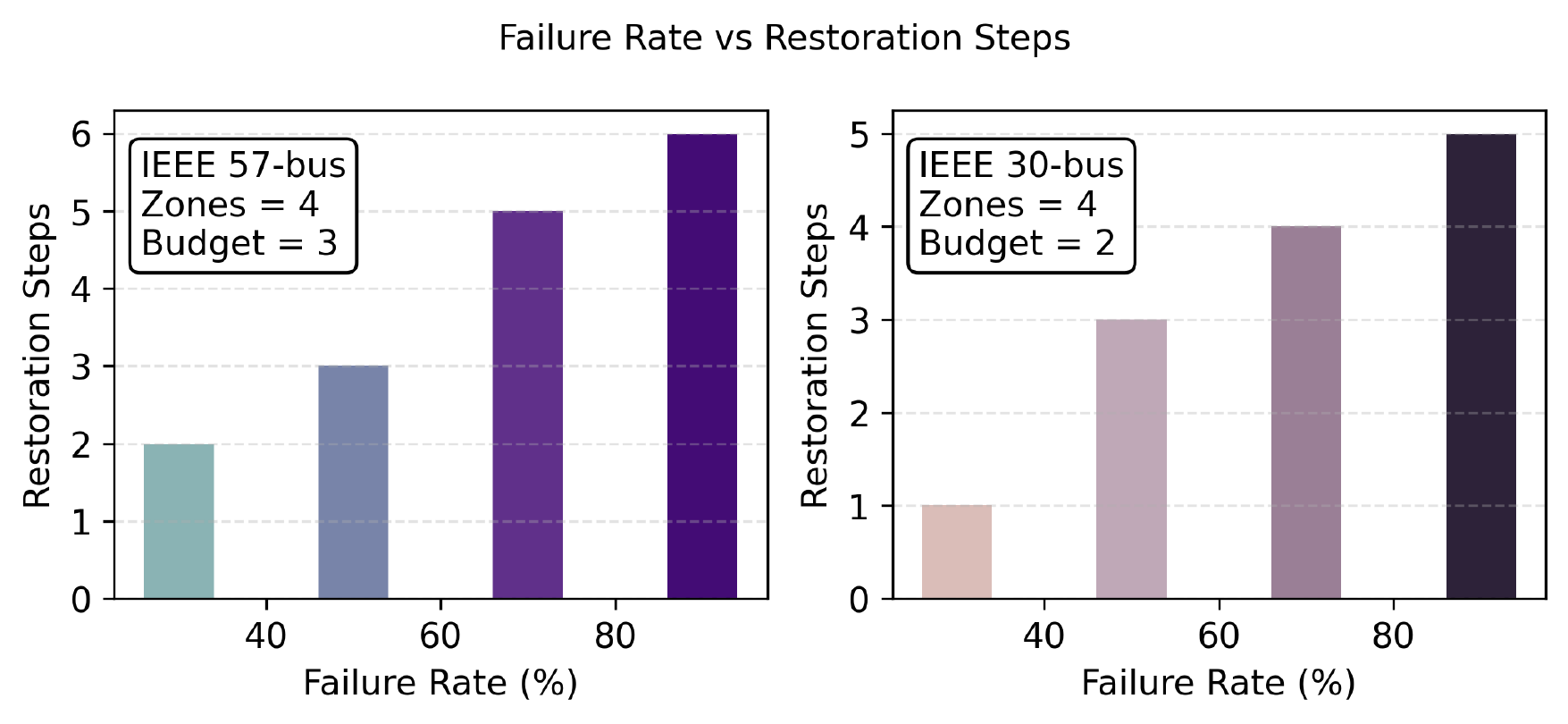}
    \caption{Restoration performance under different PMU failure rates.}
    \label{fig:fail_vs_res}
\end{figure}

Fig.~\ref{fig:fail_vs_res} shows the number of restoration steps under different failure rates for a fixed budget. As the failure rate increases, more PMUs become unavailable after the disaster, and therefore the number of required restoration steps also increases. The number of steps also depends on the system size and number of zones, since larger or more partitioned systems may require additional coordinated restoration actions to recover full observability.

\begin{table}[!t]
\centering
\footnotesize
\caption{Performance comparison with standalone MILP for 57-bus System.}

\setlength{\tabcolsep}{5pt}
\renewcommand{\arraystretch}{1.2}

\begin{tabular}{c c c c c c}
\toprule
\textbf{Number} & \textbf{Failure} & \textbf{Budget} & \textbf{Avg. Opt.} 
& \textbf{Avg. Incomp.} & \textbf{Avg.} \\
\textbf{of Zones} & \textbf{Rate} & & \textbf{Gap (\%)} 
& \textbf{Times (\%)} & \textbf{Steps} \\
\midrule

\multirow{2}{*}{1--4}
& 90\% & 1--4 & 7.76 & 21.875 & 7.93 \\
& 70\% & 1--4 & 0 & 0 & 6.43 \\

\bottomrule
\end{tabular}

\label{tab:llm_performance}
\end{table}


\begin{table*}[h]
\centering
\caption{Ablation study of framework variants.}
\label{tab:ablation}
\footnotesize
\begin{tabular}{p{5cm} p{7.5cm} p{4cm}}
\toprule
\textbf{Framework Variant} & \textbf{Performance Observations} & \textbf{Evaluation} \\
\midrule

One-shot prompt-only LLM
& Activates all failed PMUs at once and reports full observability without sequential restoration, tool-grounded computation, or explanation.
& Unreliable performance \\

Iterative LLM without tools
(manual budget allocation)
& Follows the iterative instruction and completes restoration, but lacks MILP-based decision-making and does not follow a logical restoration sequence.
& 100\% observability with 61\% more restoration steps on average \\

Iterative LLM without tools
(LLM-based budget reasoning)
& Produces partial observability improvement, but fails to complete restoration due to hallucinated or inconsistent decisions and reaches the maximum iteration limit.
& 20--30\% observability improvement \\

Proposed framework
& Uses structured tool calls, MILP-based restoration, context tracking, and explanation generation.
& 100\% observability with traceable restoration steps \\

\bottomrule
\end{tabular}
\end{table*}

\subsection{Comparative Study} This subsection compares the proposed framework with an MILP-based baseline and with different LLM variants. 

\subsubsection{Comparison with MILP-Based Baseline}

The proposed framework is compared with a deterministic MILP-based baseline that executes the same restoration functions without LLM-based orchestration. This comparison evaluates whether the LLM agent can correctly coordinate the restoration workflow. The corresponding results for the IEEE 57-bus system are reported in Table~\ref{tab:llm_performance}. When the number of zones and the restoration budget are varied from 1 to 4, the proposed framework achieves an average step gap of 7.76\% under the 90\% PMU failure rate and 0\% under the 70\% failure rate. The step gap is computed as the relative difference between the number of restoration steps required by the proposed framework and the deterministic MILP-based baseline. 
Incomplete runs and step gaps occur mainly under tight budget conditions when the system has many zones. Under the 90\% failure rate, incomplete runs are mostly observed when the system is divided into 3 or 4 zones with a very limited budget, such as 1 or 2 PMUs per step. However, it is observed that when the budget increases to 3--4 PMUs per step, the average step gap drops to 3.57\%.

In highly constrained cases, three factors contribute to incomplete runs: 1) the allocated restoration budget is insufficient; 2) the LLM stops repeated restoration when observability no longer improves under the current budget allocation; and 3) the LLM reaches its generation limit before completing the full restoration procedure. These cases indicate the need for improved context management and stronger completion handling under highly constrained restoration settings. For the IEEE 30-bus system, the proposed framework achieves 0\% step gap with a 100\% completion rate.

\subsubsection{Comparison with LLM Variants}
Table~\ref{tab:ablation} compares the proposed framework with simplified LLM variants. The one-shot prompt-only LLM is unreliable because it bypasses the sequential restoration process and reports full observability without tool-grounded verification. Iterative LLM variants improve the process but remain limited without tool-based MILP optimization, context tracking, and structured verification. The proposed framework achieves full observability while preserving tool-grounded explanations and traceability.

\section{Conclusion}
\label{sec:conclusion}

This work introduced an agentic tool-calling LLM framework for post-disaster PMU restoration and grid observability recovery. In the proposed framework, the LLM is used as a workflow orchestrator to coordinate structured tool calls, explain tool-generated restoration plans, support context-aware operator question answering, and enable human oversight before restoration actions are applied. Instead of treating restoration as a black-box optimization task, the restoration process is decomposed into modular deterministic functions, which helps reduce unsupported LLM outputs. Simulation results show that observability recovery comparable to MILP-based restoration is achieved while improved traceability, built-in explainability, and interactive operator support are provided. Future work will further demonstrate the value of the framework under diverse operating scenarios beyond static failure cases. Solutions for incomplete cases and optimality gaps will also be investigated while keeping LLM overhead minimal. This will help move the framework toward optimization-level restoration performance while reducing operators' cognitive workload. Future extensions will also explore advanced LLMs, improved context management, and multi-agent LLM frameworks.


\vspace{-2pt}

\bibliographystyle{IEEEtran}
\bibliography{myref}

@article{zhang2025grid,
  title={Grid-agent: An {LLM}-powered multi-agent system for power grid control},
  author={Zhang, Yan and Saber, Ahmad Mohammad and Youssef, Amr and Kundur, Deepa},
  journal={arXiv preprint arXiv:2508.05702},
  year={2025}
}

@inproceedings{jin2025gridmind,
  title={{GridMind}: {LLMs}-powered agents for power system analysis and operations},
  author={Jin, Hongwei and Kim, Kibaek and Kwon, Jonghwan},
  booktitle={Proceedings of the SC'25 Workshops of the International Conference for High Performance Computing, Networking, Storage and Analysis},
  pages={560--568},
  year={2025}
}

@ARTICLE{agenticAI,
  author={Acharya, Deepak Bhaskar and Kuppan, Karthigeyan and Divya, B.},
  journal={IEEE Access}, 
  title={Agentic {AI}: Autonomous Intelligence for Complex Goals—A Comprehensive Survey}, 
  year={2025},
  volume={13},
  number={},
  pages={18912-18936},
  doi={10.1109/ACCESS.2025.3532853}}

@article{agenticAI2,
  title={Agentic {AI} Systems in Electrical Power Systems Engineering: Current State-of-the-Art and Challenges},
  author={Ghosh, Soham and Mittal, Gaurav},
  journal={arXiv preprint arXiv:2511.14478},
  year={2025}
}

@article{phadke2018pmu,
  author  = {Phadke, A. G. and Bi, T.},
  title   = {Phasor Measurement Units, {WAMS}, and Their Applications in Protection and Control of Power Systems},
  journal = {Journal of Modern Power Systems and Clean Energy},
  year    = {2018},
  volume  = {6},
  number  = {4},
  pages   = {619--629},
  month   = jul,
  doi     = {10.1007/s40565-018-0423-3}
}

@article{baldwin1993pmu,
  author  = {Baldwin, T. L. and Mili, L. and Boisen, M. B. and Adapa, R.},
  title   = {Power System Observability with Minimal Phasor Measurement Placement},
  journal = {IEEE Transactions on Power Systems},
  year    = {1993},
  volume  = {8},
  number  = {2},
  pages   = {707--715},
  month   = may,
  doi     = {10.1109/59.260810}
}

@article{liang2017fdi,
  author  = {Liang, G. and Zhao, J. and Luo, F. and Weller, S. R. and Dong, Z. Y.},
  title   = {A Review of False Data Injection Attacks Against Modern Power Systems},
  journal = {IEEE Transactions on Smart Grid},
  year    = {2017},
  volume  = {8},
  number  = {4},
  pages   = {1630--1638},
  month   = jul,
  doi     = {10.1109/TSG.2015.2495133}
}

@article{lei2019resilient,
  author  = {Lei, S. and Chen, C. and Li, Y. and Hou, Y.},
  title   = {Resilient Disaster Recovery Logistics of Distribution Systems: Co-Optimize Service Restoration with Repair Crew and Mobile Power Source Dispatch},
  journal = {IEEE Transactions on Smart Grid},
  year    = {2019},
  volume  = {10},
  number  = {6},
  pages   = {6187--6202},
  month   = nov,
  doi     = {10.1109/TSG.2019.2899353}
}

@article{edib2023cyber,
  author  = {Edib, S. N. and Lin, Y. and Vokkarane, V. M. and Qiu, F. and Yao, R. and Chen, B.},
  title   = {Cyber Restoration of Power Systems: Concept and Methodology for Resilient Observability},
  journal = {IEEE Transactions on Systems, Man, and Cybernetics: Systems},
  year    = {2023},
  volume  = {53},
  number  = {8},
  pages   = {4773--4786},
  month   = aug,
  doi     = {10.1109/TSMC.2023.3258412}
}

@article{edib2021optimal,
  author  = {Edib, S. N. and Lin, Y. and Vokkarane, V. M. and Qiu, F. and Yao, R. and Zhao, D.},
  title   = {Optimal {PMU} Restoration for Power System Observability Recovery After Massive Attacks},
  journal = {IEEE Transactions on Smart Grid},
  year    = {2021},
  volume  = {12},
  number  = {2},
  pages   = {1565--1576},
  month   = mar,
  doi     = {10.1109/TSG.2020.3028761}
}

@inproceedings{haggi2020multi,
  author    = {Haggi, H. and Sun, W.},
  title     = {Multi-Objective {PMU} Allocation for Resilient Power System Monitoring},
  booktitle = {2020 IEEE Power and Energy Society General Meeting (PESGM)},
  year      = {2020},
  pages     = {1--5},
  month     = aug,
  publisher = {IEEE},
  address   = {Montreal, QC, Canada},
  doi       = {10.1109/PESGM41954.2020.9281963}
}

@misc{openai_gpt4,
  title        = {GPT-4 Technical Report},
  author       = {{OpenAI}},
  year         = {2023},
  eprint       = {2303.08774},
  archivePrefix= {arXiv},
  primaryClass = {cs.CL}
}

@article{
reason_fail,

  title={Large Language Model Reasoning Failures},
  author={Song, Peiyang and Han, Pengrui and Goodman, Noah},
  journal={arXiv preprint arXiv:2602.06176},
  year={2026}
}

@misc{github_repo,
  author       = {{SIU Power Lab}},
  title        = {{LLM} Tool Example},
  year         = {2026},
  howpublished = {\url{https://github.com/siu-power-lab/llm_tool_example/tree/main}},
  note         = {Accessed: 2026-05-03}
}

\end{document}